\documentclass[twocolumn,twosidea4paper,12pt]{jtcs}
\usepackage{bm}
\usepackage{amssymb}
\usepackage{graphicx}

\begin{document}

\title{Bayesian Analysis of $C_{x'}$ and $C_{z'}$ Double Polarizations in Kaon Photoproduction}

\author{P. T. P. Hutauruk}
\address{School of Physics and Astronomy, University of Glasgow, Scotland-United Kingdom G12 8QQ}

\date{Submitted on, \today}

\abstract{Have been analyzed the latest experimental data for $\gamma + p \to K^{+} + \Lambda$ reaction of $C_{x'}$ and $C_{z'}$ double polarizations \cite{bradford07}. In theoretical calculation, all of these observables can be classified into four Legendre classes and represented by associated Legendre polynomial function itself \cite{fasano92}. In this analysis we attempt to determine the best data model for both observables. We use the bayesian technique to select the best model by calculating the posterior probabilities and comparing the posterior among the models. The posteriors probabilities for each data model are computed using a Nested sampling integration \cite{skilling06}. From this analysis we concluded that $C_{x'}$ and $C_{z'}$ double polarizations require two and three order of associated Legendre polynomials respectively to describe the data well. The extracted coefficients of each observable will also be presented. It shows the structure of baryon resonances qualitatively.}

\email{hparada@gfti.fisika.net}

\maketitle

\section{Introduction}
The world database for the reaction $\gamma + p \to K^{+} + \Lambda$ is more available now. This gives a possibility to analyze the data more accurately. Recently the newest experimental data for the $K^{+}\Lambda$ channel of photon asymmetry ($\Sigma$), target polarization (T), recoil polarization (P), and $O_{x'}$ and $O_{z'}$ double polarizations has been collected yet \cite{craig08}. Unfortunately all of the experimental data do not published yet. Additionally the experimental data of $C_{x}$ and $C_{z}$ double polarizations has been published since several years ago. Therefore in this work we only focus analyzing the data of these both double observables. Furthermore the experimental data for other observables such as G asymmetry will available soon from Jefferson Lab. Newport News, USA.

In this work we try to extract the information from the experimental data directly in order to understand about baryon resonances for $K\Lambda$ channel, which become a longstanding problem in this field. We analyze the data using associated Legendre polynomials which represented 16 observables in the $K^{+}\Lambda$ \cite{fasano92}. Based upon the Legendre classes, ${\cal L}_0(\hat{{\bf I}};\hat{{\bf E}};\hat{{\bf C_{z'}}};\hat{{\bf L_{z'}}})$, ${\cal L}_{1a}(\hat{{\bf P}}; \hat{{\bf H}}; \hat{{\bf C_{x'}}}; \hat{{\bf L_{x'}}})$, ${\cal L}_{1b}(\hat{{\bf T}}; \hat{{\bf F}}; \hat{{\bf O_{x'}}};\hat{{\bf T_{z'}}})$ and ${\cal L}_2(\hat{{\bf {\Sigma}}}; \hat{{\bf G}}; \hat{{\bf O_{z'}}}; \hat{{\bf T_{x'}}})$, we construct the data model for each observable. In general we have two steps to determine the best model. Firstly to compute the maximum posterior for each photon energy for each observable. Secondly to compare the maximum posterior among the models with different order of Legendre polynomials. We then employ the Nested sampling integration to execute the multi-integral over the parameters of Legendre polynomials. Nested sampling is a very powerful technique to evaluate the multi-dimensions integral into one dimension integral. It also has been used in many fields such as astrophysics \cite{pia06}, cosmology and high energy physics.

To compare different data models by evaluating the ratios of the maximum posterior, given a set of experimental data ${\bf D}$. Using Bayes'theorem, the ratio $\mathcal{R}$ of $M_i$ and $M_0$ data models can be written as follows: 
\begin{eqnarray}
\mathcal{R} &=& \frac{p(M_{i}\mid {\bf D})}{p(M_{0}\mid {\bf D})},\nonumber \\
  &=& \frac{p({\bf D}\mid M_{i})}{p({\bf D}\mid M_{0})} \times \frac{p(M_{i})}{p(M_{0})}.
\label{Bayes}
\end{eqnarray}
Where $p(M_{i}\mid {\bf D})$ is maximum posterior for the $M_{i}$ data models, $p({\bf D}\mid M_{i})$ is probability data would be obtained, assuming $M_{i}$ to be true where i= 1,2,3...L. With no pior prejudice as to which is correct, we obtain the ratio of the likelihoods:
\begin{eqnarray}
\mathcal{R} &=& \frac{p({\bf D}\mid M_{i})}{p({\bf D} \mid M_{0})}.
\label{likelihoodratio}
\end{eqnarray}
The likelihood $p({\bf D}\mid M_{i})$ is an integral over the joint likelihood $p({\bf D},\{A_{i}\}\mid M_{i})$ or Bayesian evidence (${\bf \mathcal{Z}}$), where $\{A_{i}\}$ represents a set of free parameters:
\begin{eqnarray}
p({\bf D}\mid M_{i}) &=& \int ...\int p({\bf D},\{A_{i}\}\mid M_{i}) d^{i} A_{i}, \nonumber \\
                     &=& \int ...\int p({\bf D}\mid \{A_{i}\}, M_{i}), \nonumber \\
                     &\times& p(\{A_{i}\}\mid M_{i}) d^{i}A_{i}.
\label{likeli2}
\end{eqnarray}
For simplicity the Eq.(\ref{likeli2}) can also be written as follows:
\begin{eqnarray}
{\bf \mathcal{Z}} &=& \int {\mathcal{L}}({\bf A_{i}}) \pi({\bf A_{i}}) d{\bf A_{i}}.
\label{likeli3}
\end{eqnarray}
Where ${\mathcal{L}}({\bf A_{i}})$ is the likelihood functions and $\pi({\bf A_{i}})$ is the prior distribution.

Nested sampling is a Monte Carlo integration technique for evaluating the integral of a likelihood function or Bayesian Evidence in Eq.(\ref{likeli3}) over its range of parameters which is developed by Skilling \cite{skilling06}. This tehnique exploits the relation between the likelihood and prior volume to transform the multidimensional of evidence integral into one dimensional integral. The prior volume X is defined by $dX = \pi({\Theta}) d^{D}_{\Theta}$, so that:
\begin{eqnarray}
X(\lambda) = \int_{{\mathcal{L}}(\Theta)} \pi(\Theta) d^{D}_{\Theta},
\label{prior}
\end{eqnarray}
where the integral extend over the region of the parameter space contained the iso-likelihood contour ${\mathcal{L}}(\Theta)= \lambda$. Asumming that ${\mathcal{L}}(X)$ is monotonically decreasing function of X which is trivially satisfied for most posteriors. The evidence integral can be written as follows:
\begin{eqnarray}
\mathcal{Z} = \int ^{1}_{0} {\mathcal{L}}(X) dX.
\label{evidence}
\end{eqnarray}
A more detailed of this technique can be found in Ref.\cite{skilling06}. 
     
\section{Results}
The experimental data available for $K\Lambda$ of $C_{x}$ dan $C_{z}$ double polarization measured at unprime coordinat system, whereas in our analysis we require the data of $C_{x'}$ and $C_{z'}$ double polarization which were measured at prime coordinate system (outgoing kaon or z'-axis). Hence we have to transform these observables using a standard rotation matrix as follows:
\begin{eqnarray}
C_{x'} = C_{x} \cos{\theta}- C_{z}\sin{\theta},\nonumber \\
C_{z'} = C_{x} \sin{\theta}+ C_{z}\cos{\theta}.
\label{transform}
\end{eqnarray}   
Where $\theta$ is the kaon scattering angle. Then the results of this transformation will be used in the analysis.

\begin{figure}[t]
        \centering 
	\includegraphics[width=8cm]{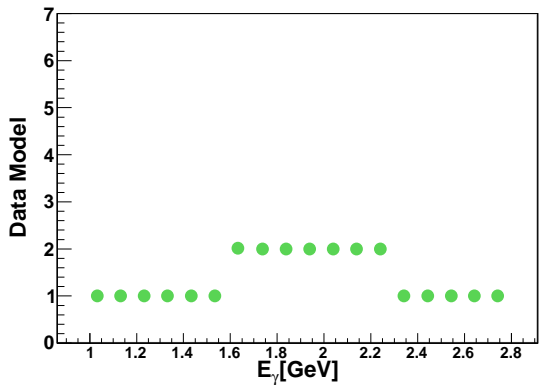}
	\caption{The best data model for each photon energy for $C_{x'}$ double polarization.} 
	\label{bestmodelcx}
\end{figure}

\begin{figure*}[t]
        \centering 
	\includegraphics[width=15cm]{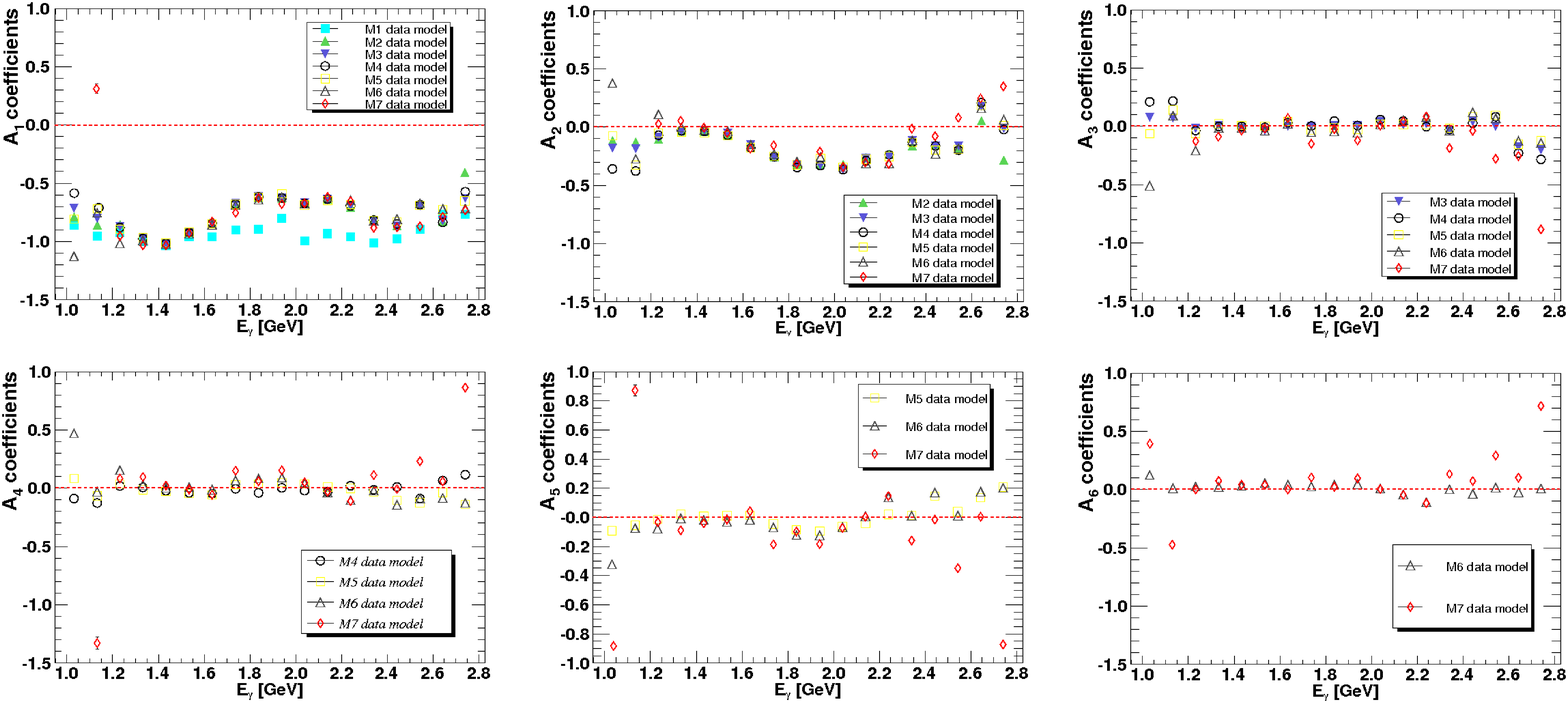}
	\caption{Extracted coefficients of $C_{x'}$ double polarization for each photon energies.} 
	\label{extractcx}
\end{figure*}

Using Eq.(\ref{evidence}) we computed the evidence for each photon energies ($E_{\gamma}$). For calculating the posterior we chosen the uniform prior distribution $\pi(\Theta)$. We then compared the posterior of the data model with different order of associated Legendre polynomilas for each photon energies. The best data model for each photon energy of $C_{x'}$ double polarization are shown in Fig.\ref{bestmodelcx}. Fig.\ref{bestmodelcx} provides most experimental data of $C_{x'}$ double polarization can be described well by $M_{1}$ data model. Extracted coefficients of the best model for $C_{x'}$ double polarization provided in Fig.\ref{extractcx}. The extracted coefficients results for other models are also shown in Fig.\ref{extractcx}. 

\begin{figure}[t]
        \centering 
	\includegraphics[width=8cm]{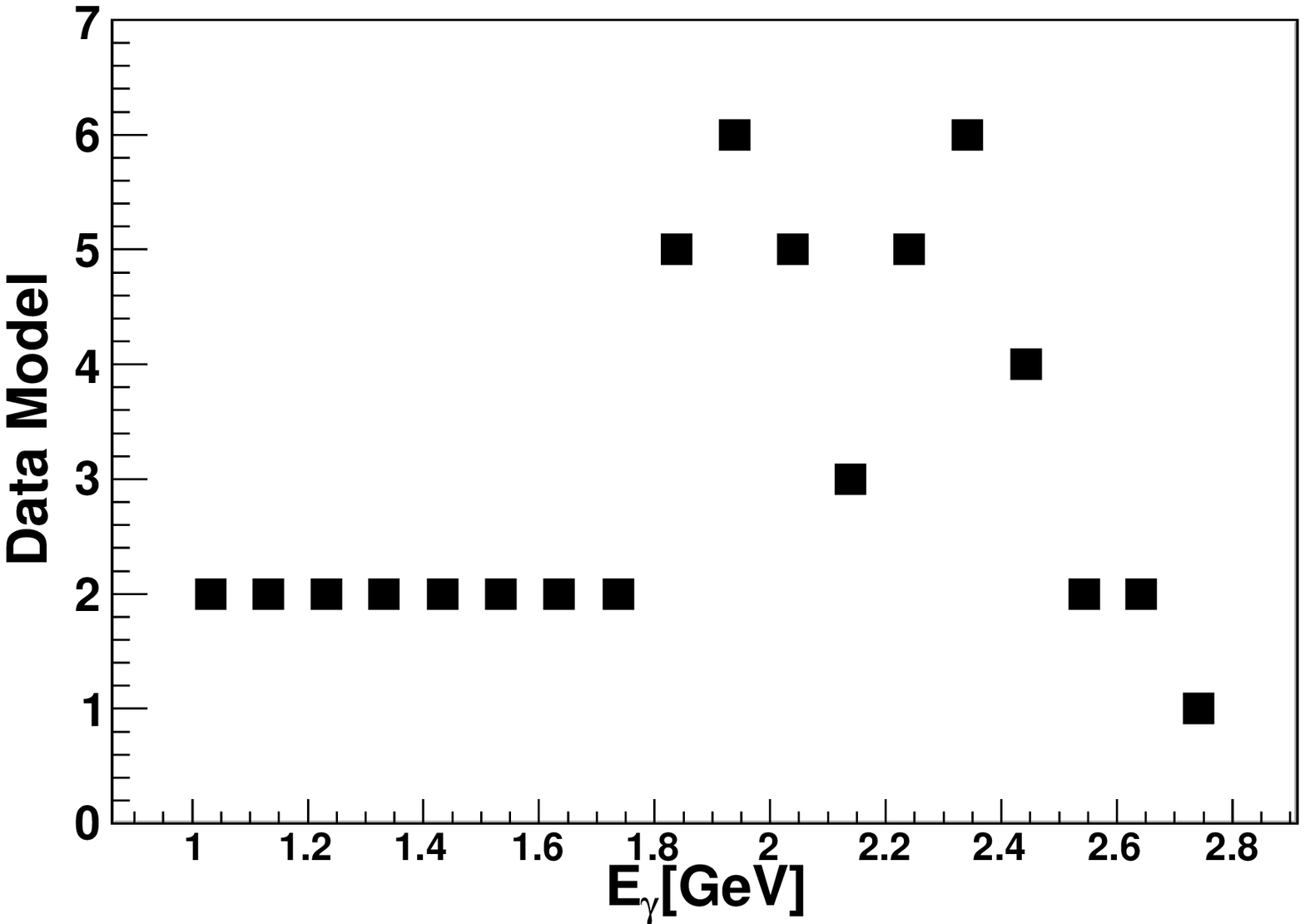}
	\caption{The best data model for each photon energy for $C_{z'}$ double polarization.} 
	\label{bestmodelcz}
\end{figure}

\begin{figure*}[t]
        \centering 
	\includegraphics[width=15cm]{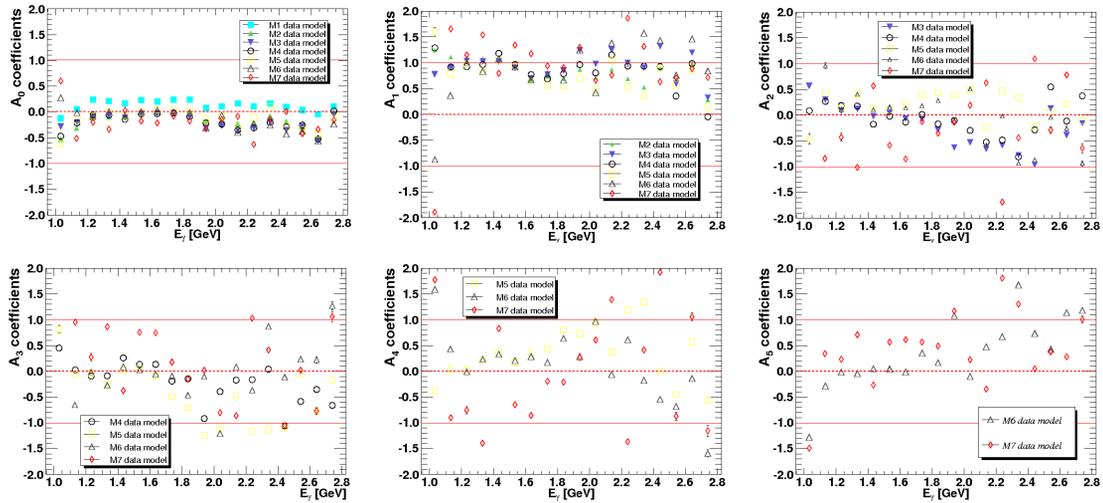}
	\caption{Extracted coefficients of $C_{z'}$ double polarization for each photon energies.} 
	\label{extractcz}
\end{figure*}

With similar procedures, the best data model for $C_{z'}$ double polarization are shown in Fig.\ref{bestmodelcz}. The most experimental data for $C_{z'}$ double polarization can described by $M_{2}$ data model. The extracted coefficients for this observale provided in Fig.\ref{extractcz}. Generally the extracted coefficients results of the best data model for $C_{x'}$ and $C_{z'}$ double polarizations may reveal the baryon resonances. 
\section{Conclusion}
Bayesian analysis is a powerful tools for determining the best data model. We have analyzed the associated Legendre polynomials of $C_{x'}$ and $C_{z'}$ double polarizations data. We computed the evidence of the data models by using the Nested sampling integration. From this analysis we found that $C_{x'}$ and $C_{z'}$ double polarizations require two and three order of associated Legendre polynomials respectively. 

\section*{Acknowledgments}
This work was supported by the Scottish Universities Physics Alliance (SUPA) Fellowship.

\end{document}